\documentclass{webofc}
\usepackage[varg]{txfonts}   

\usepackage{wrapfig}
\usepackage{hyperref}

\usepackage{caption}

\begin{document}
\title{A truly cylindrical inner tracker for ALICE}
%
%

\author{\firstname{Alperen} \lastname{Y\"unc\"u}\inst{1}\fnsep\thanks{\email{alperen.yuncu@cern.ch}}}

\institute{Ruprecht Karl University of Heidelberg}

\abstract{%
After the successful installation and first operation of the upgraded Inner Tracking System (ITS2), which consists of about $10\,$m$^2$ of monolithic silicon pixel sensors, ALICE is pioneering the usage of bent, wafer-scale pixel sensors for the ITS3 for LHC Run 4.
Sensors larger than typical reticle sizes can be produced using the technique of stitching.
At thicknesses of about $30\,\mu$m, silicon is flexible enough to be bent to radii of the order of $1\,$cm. 
By cooling such sensors with a forced air flow, it becomes possible to construct truly cylindrical layers which consist practically only of the silicon sensors.
The reduction of the material budget and the improved pointing resolution will allow new measurements, in particular of heavy-flavour decays and electromagnetic probes.
In this presentation, we will report on the sensor developments, the performance of bent sensors in test beams, and the mechanical studies on truly cylindrical layers.
}
\maketitle
\section{Introduction}

\begin{figure}[h]
\begin{minipage}{0.54\linewidth}
ALICE is an experiment at the CERN LHC mainly dedicated at studying heavy-ion collusions.
Thus, ALICE is optimised for large particle multiplicities, and to measure particles with low momenta.
The Inner Tracking System is the detector closest to the interaction point.
The current Inner Tracking System (ITS2), installed in the ALICE Experiment,  is ultra light (radiation length of $0.35\%\,$X$_0$ per layer) but rather densely packed \cite{ITS2}.
The material budget distribution of the innermost layer of the ITS2 is given in Fig.~\ref{mbudget}.
Only 14\% of the total contribution to the material budget is from the silicon sensors.
The rest consists of mechanical support structures (carbon and glue), circuitry for power distribution and data transfer (aluminum and kapton), and cooling system (water).
Removing all the material except silicon sensors which are thinned down to $20$--$40\,\mu$m would reduce the budget to $0.05\%\,$X$_0$ per layer.
\end{minipage}\quad%
\begin{minipage}{0.4\linewidth}
    \includegraphics[width=\textwidth,clip]{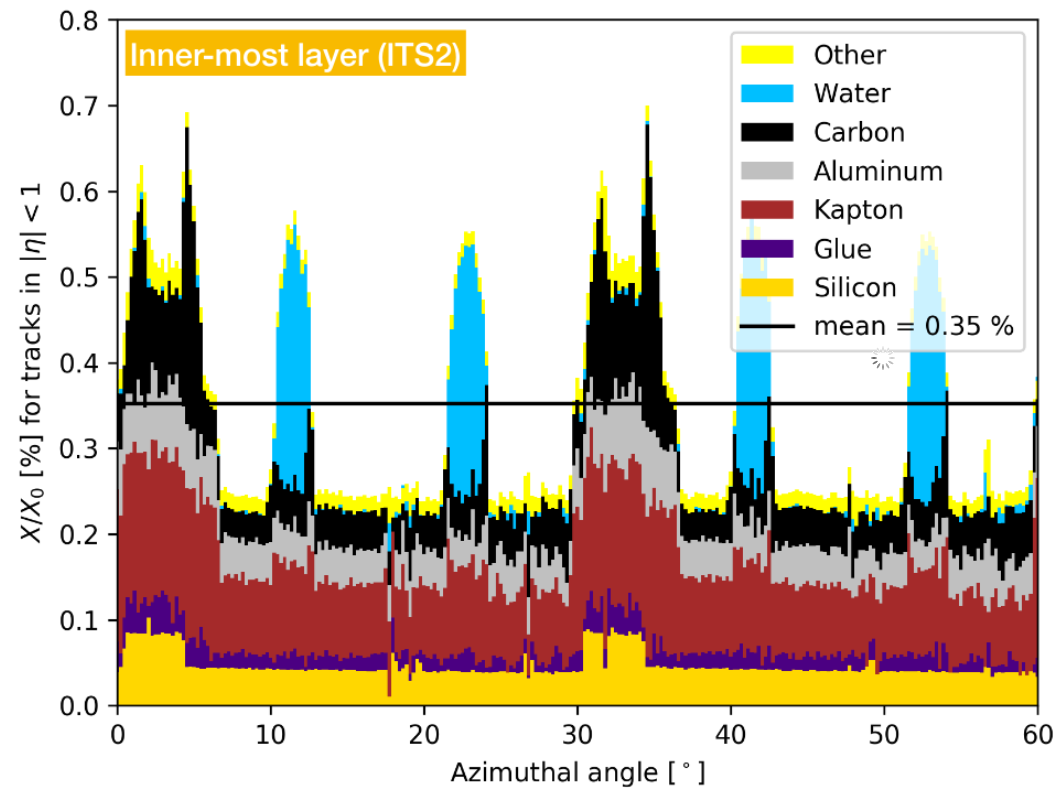}
    \captionof{figure}{The material budget of ITS2 \cite{perform}.}
\label{mbudget}       
\end{minipage}  %
\end{figure}
\clearpage
\begin{wrapfigure}{r}{0.50\textwidth}
\centering
    \includegraphics[width=0.5\textwidth,clip]{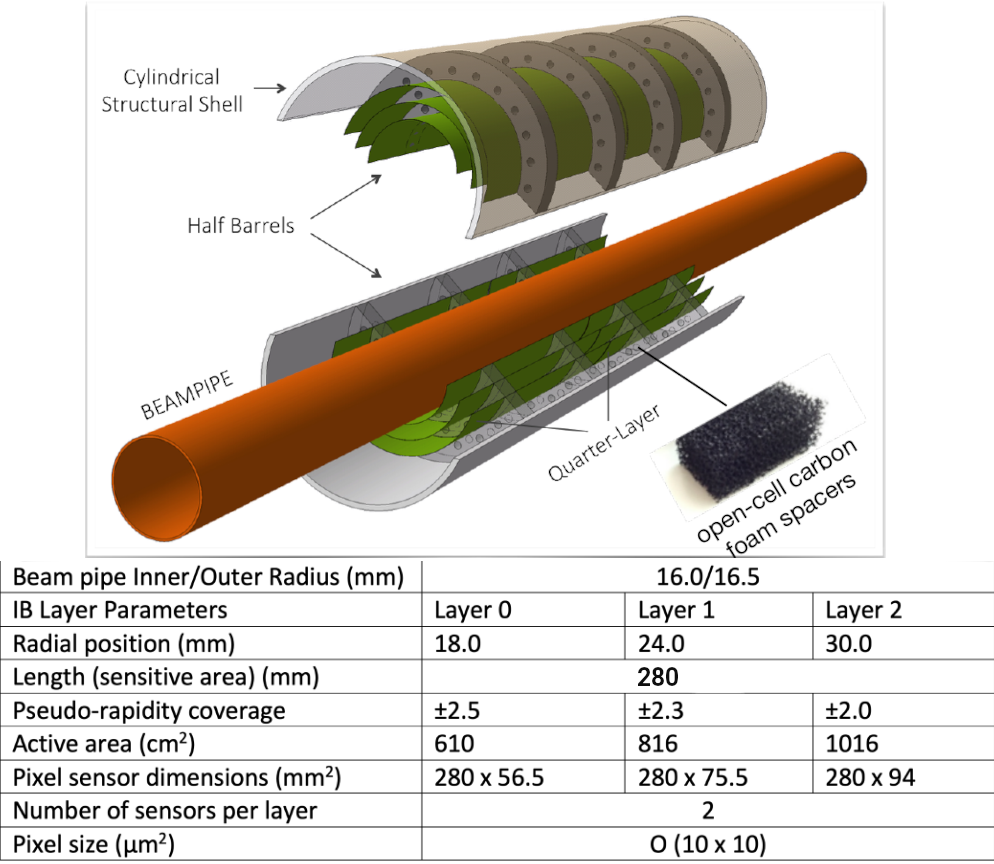}
\caption{ITS3 Layout}
\label{consept}       
\end{wrapfigure}

\section{ITS3 Project}
The inner barrel of the inner tracking system will be replaced by ITS3 in LHC Long Shutdown 3 \cite{perform}.
ITS3 will consist of 6 truly (half-)cylindrical, wafer scale sensors  rather than 432 small flat sensors, as shown in Fig.~\ref{consept}.
Each sensor will have $280\,$mm length and will be fabricated using stitching of CMOS Imaging Sensors (CIS) produced in a $65\,$nm process.
The sensors will be bent to $18$, $24$, and $30\,$mm curvature radii to form two half-barrels, each comprising three half-layers, that will be mounted around the beam pipe.
The industrial procedure called \textit{stitching} helps to propagate the power and signal through ASIC blocks which makes it possible to remove all circuitry inside the detector.
The bent shape provides additional mechanical stabilty so that the wafer-scale sensor can be held by carbon foams ribs only.

\subsection{Expected Performance}
\begin{figure}[h]
\centering
\includegraphics[width=0.45\linewidth]{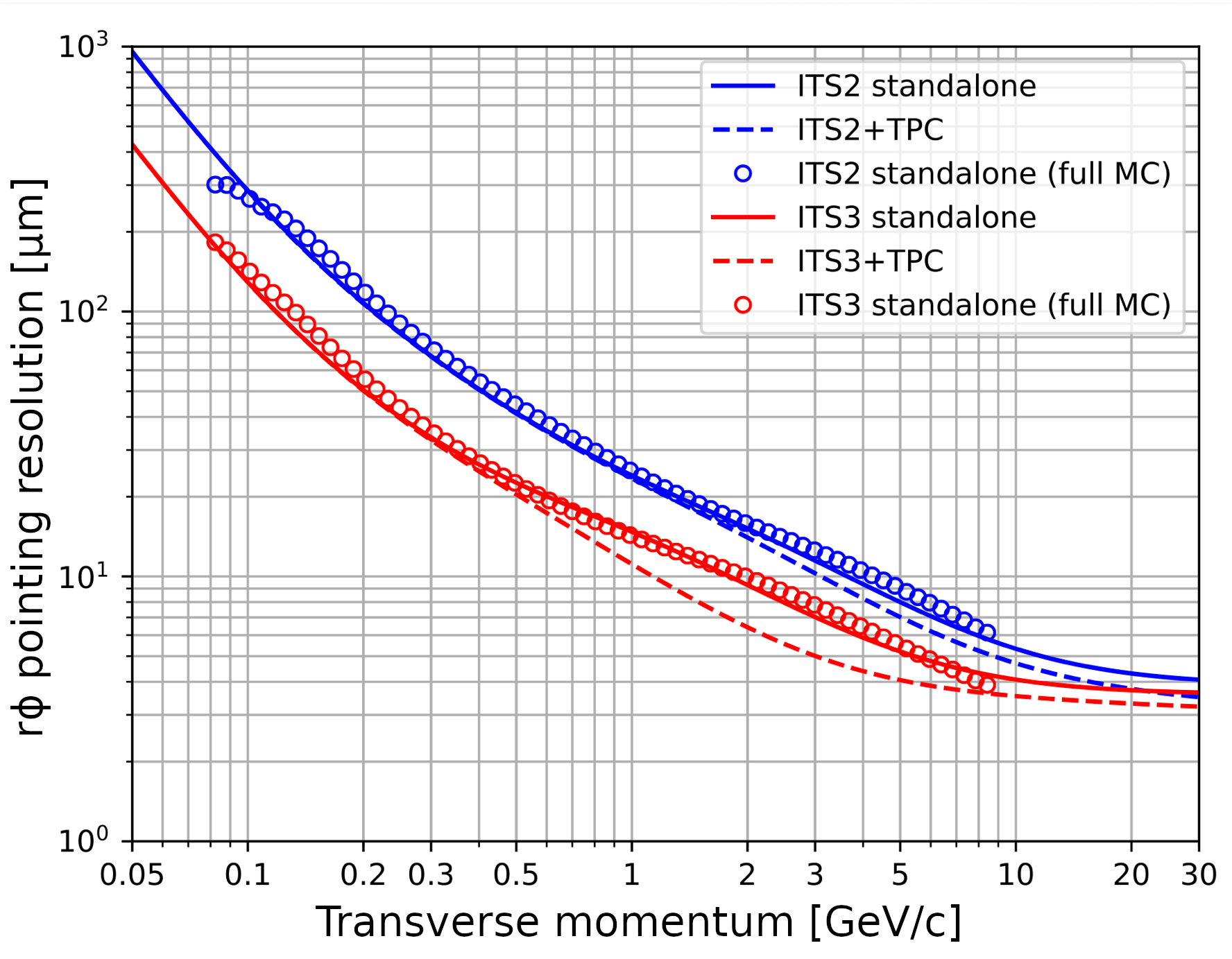}%
\includegraphics[width=0.45\linewidth]{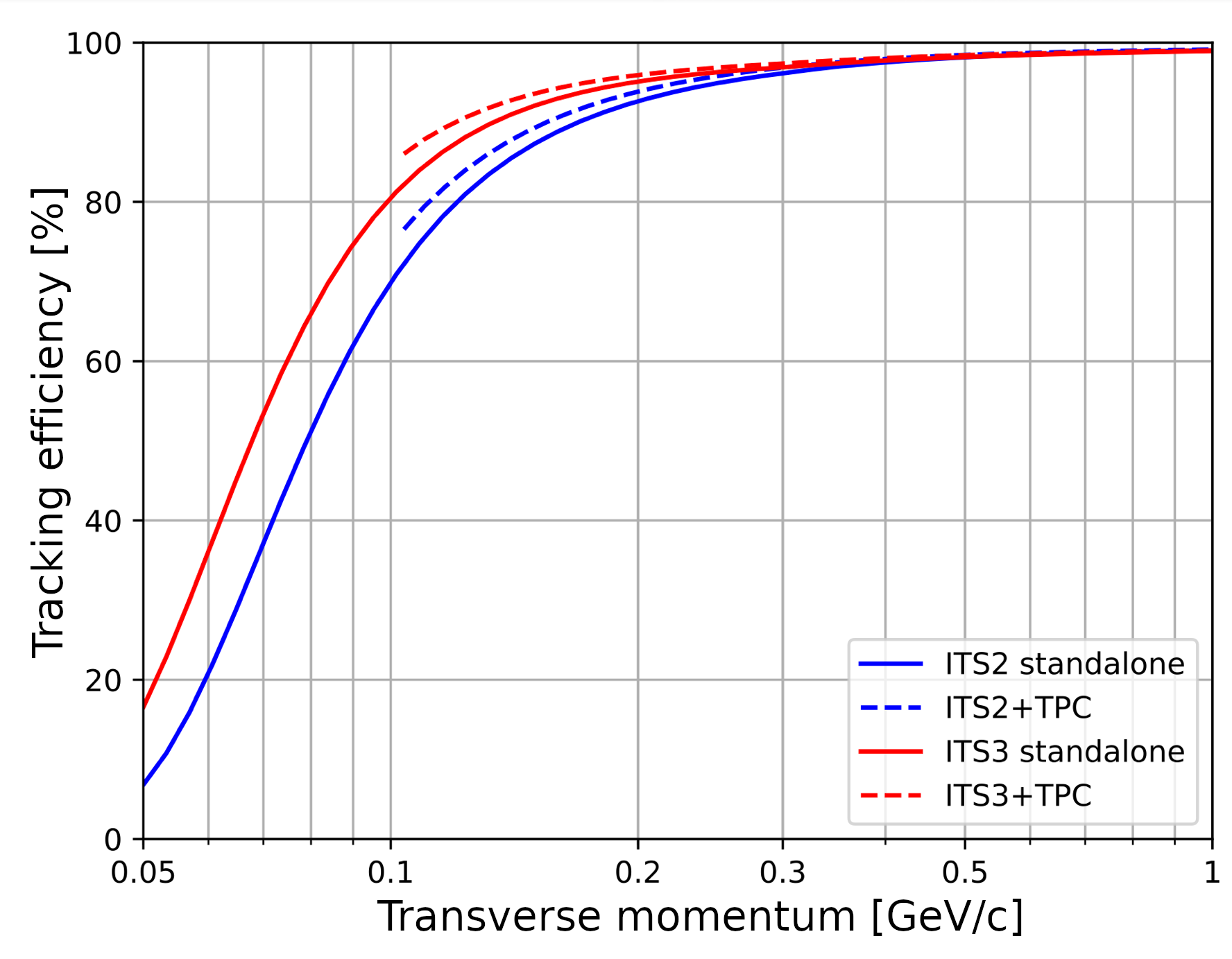}%
    \caption{Impact parameter resolution (left) and track finding efficiency (right) as a function of transverse momentum \cite{perform}.}
    \label{perform}       
\end{figure}

The projected impact parameter resolution and tracking efficiency are shown in Fig.~\ref{perform}.
The impact parameter resolution is improved by a factor of two with respect to the ITS2 performance for transverse momenta up to $5\,$GeV/c.
A significant improvement in the track finding efficiency occurs, especially, for transverse momenta lower than $200\,$MeV/c thanks to the reduced material budget.

\clearpage

\begin{figure}[h]
\begin{minipage}{0.44\linewidth}
\centering
    \includegraphics[width=\textwidth,clip]{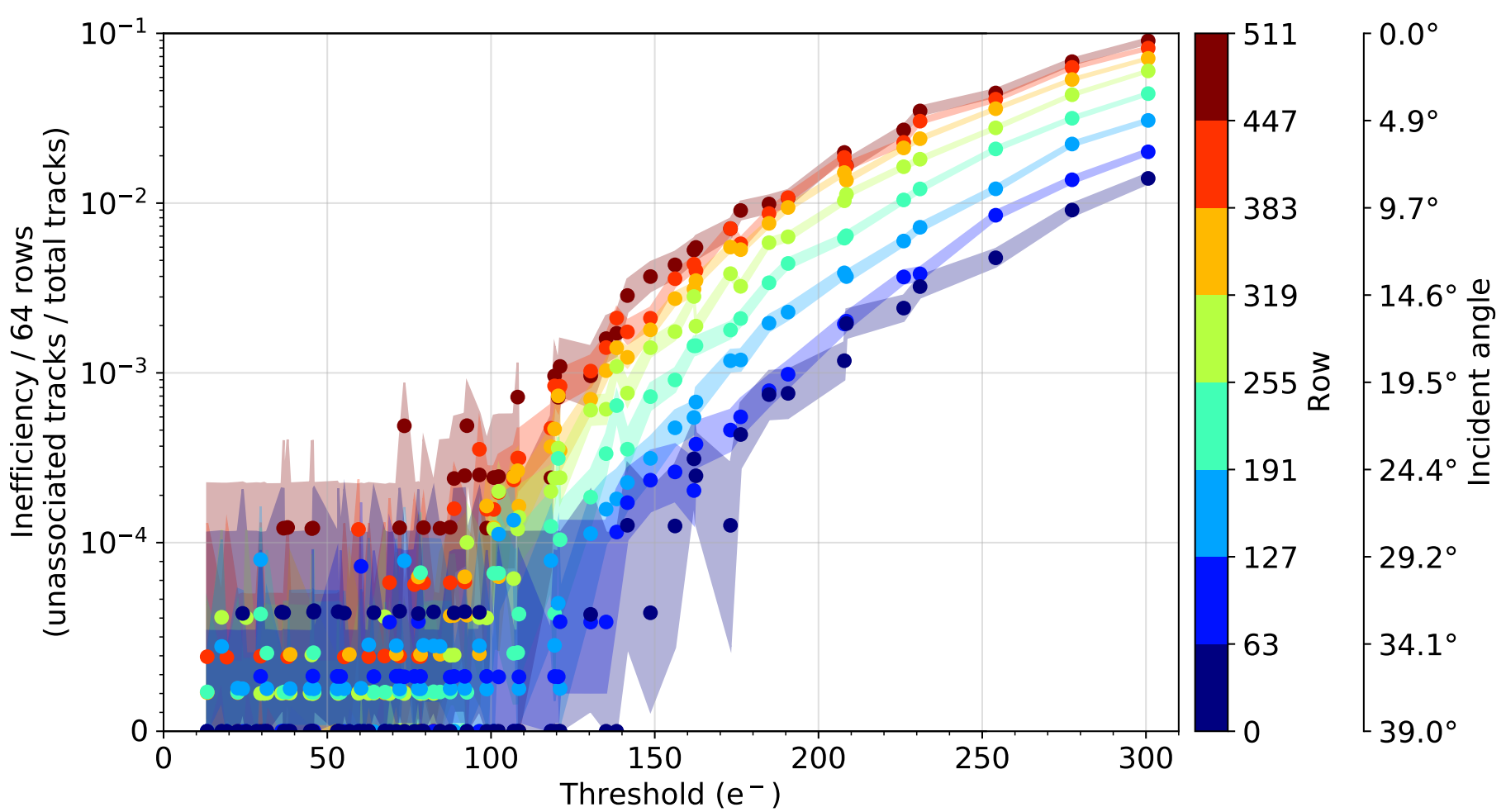}
    \captionof{figure}{Inefficiency as a function of threshold for different rows and incident angles. Each data point corresponds to at least 8k tracks.\cite{testbeam}}
\label{testbeam}       
\end{minipage} \quad %
\begin{minipage}{0.5\linewidth}
\section{Bent Sensors}
Bending tests show that even for the current thickness ($50\,\mu$m) it is feasable to reach the target radii without breaking the sensor.
Since the flexibility is inversely propotional to the third power of the thickness, it is safer and easier to achieve the same curvature with thinner chips.
The performance of the bent ASICs has been investigated with test beams.
As shown in Fig.~\ref{testbeam}, the efficiency of bent ASICs is very good. \hspace*{\fill} \break
\end{minipage}%
\end{figure}

\begin{figure}[h]
\begin{minipage}{0.64\linewidth}
\section{Super-ALPIDE}
The Super ALPIDE is a series of arrays of ALPIDE chips with thickness of $30$, $40$, and $50\,\mu$m.
They are an intermediate step from the ALPIDE chips, which are used in ITS2, towards wafer-scale sensors, which will be used in ITS3.
    Since ultra-thin sensors have not yet been produced in targeted sizes, simple tasks (like removing chips from wafer or how to transport and how to store them) have required investigations.
Thus, significant know-how is obtained about how to handle ultra thin and very large sensors.
Moreover, they will be used to investigate how to perform mechanical mounting of large scale bent sensors.\hspace*{\fill} \break
\end{minipage}%
\begin{minipage}{0.34\linewidth}
    \centering
    \includegraphics[width=\textwidth,clip]{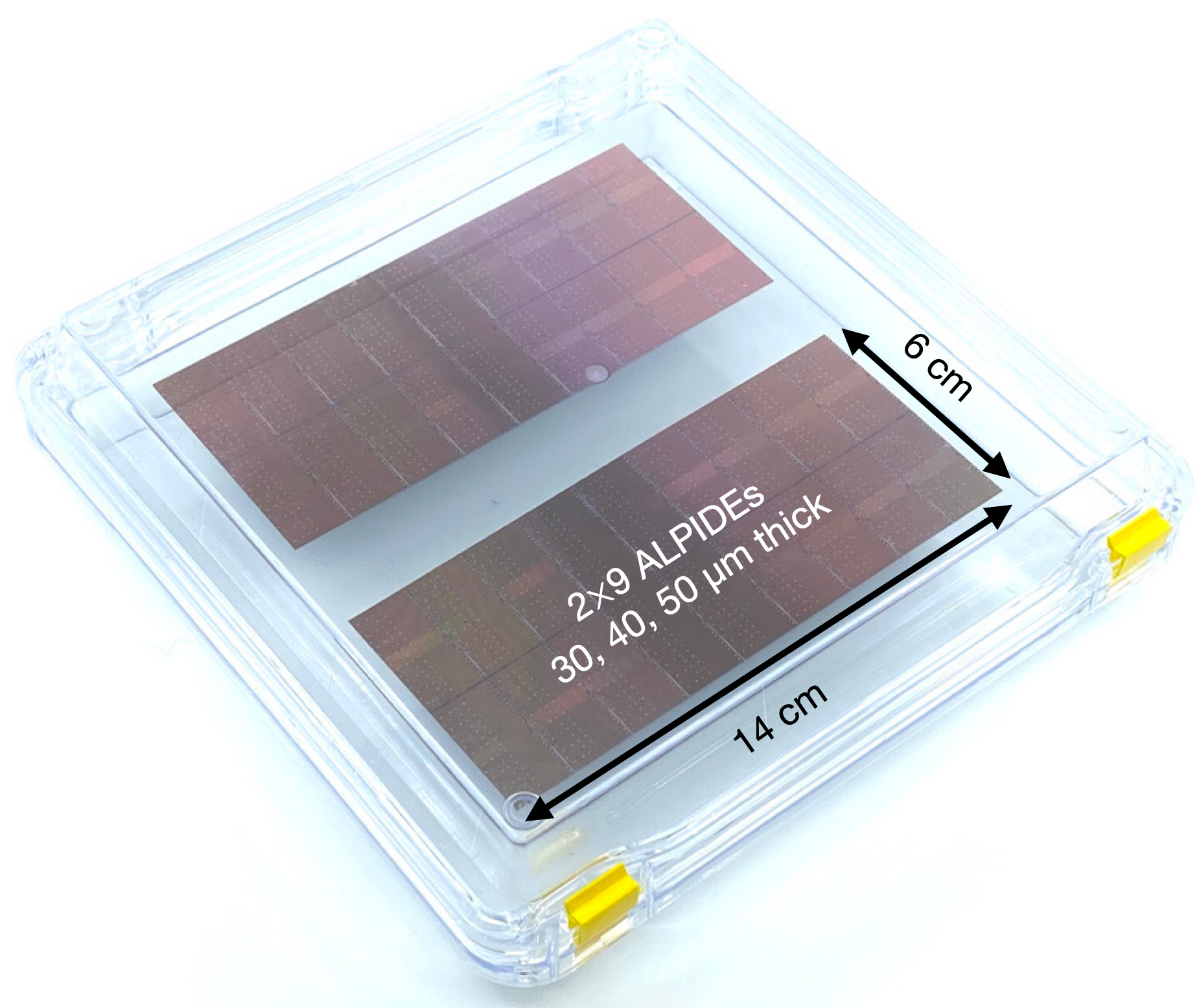}
    \captionof{figure}{Super ALPIDE}
\end{minipage}%
\end{figure}

\begin{figure*}[h]
\centering
    \includegraphics[width=0.8\textwidth,clip]{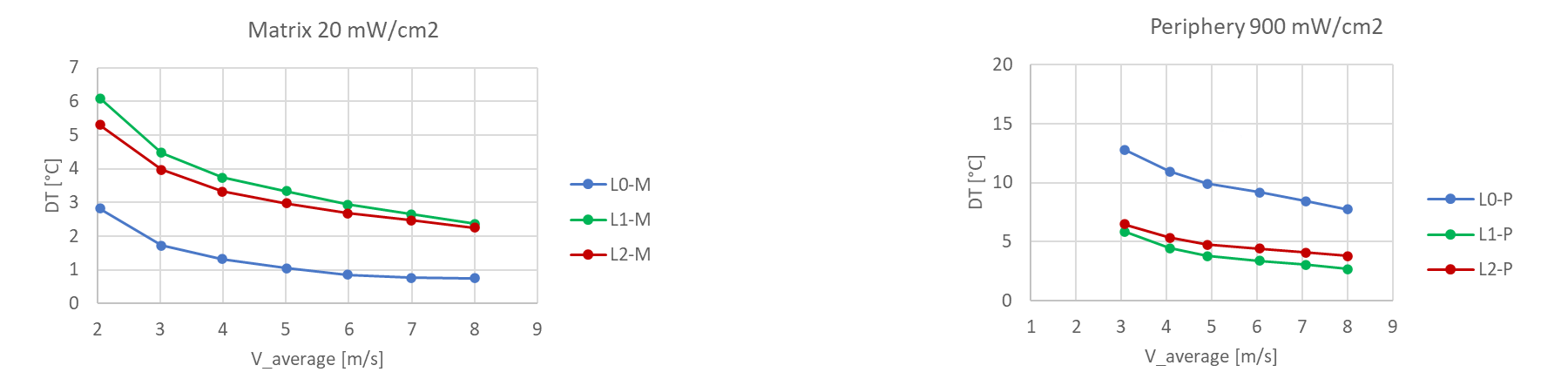}
    \caption{Temperature changes relative to the ambient temperature measured in the wind tunnel as a function of air speed}
\label{cool}       
\end{figure*}
\section{Cooling}
The water cooling system will be removed in ITS3.
Instead of water cooling, air cooling will be used.
Wind tunnel tests have been performed using a full scale model including heaters to emulate a power dissipation of $20\,$mW/cm$^2$ for the pixel matrix and $900\,$mW/cm$^2$ for the periphery where the read-out electronics is located.
Fig.~\ref{cool} shows the difference in temperature of the sensors and ambience as a function of the air flow velocity in a wind tunnel test. 
It is found that $6\,$m/s air speed is enough to keep the temperature difference lower than $4^\circ\,$C for the pixel matrix and lower than $10^\circ\,$C for the periphery.

\section{ASIC Development}
The industrial stitching technique is used in ITS3.
The stitching is used to provide power and signal connections to the repeated sensor units.
The stitching is the repeated and aligned exposure of the same mask\, which creates a large periodical and interconnected structure that can span over a whole wafer.
Metal strips distributing supply voltages traverse the stitching boundary.
Wiring through the boundaries is also used to implement long range on-chip interconnect busses for the slow control and data readout. 
Hence, wafer scale sensors are possible with stitching.
In addition, the stitching makes it possible to remove all external circuitry in the detector active area and to embed it in the sensor chips.
\begin{figure}[h]
    \centering
    \includegraphics[width=0.8\textwidth]{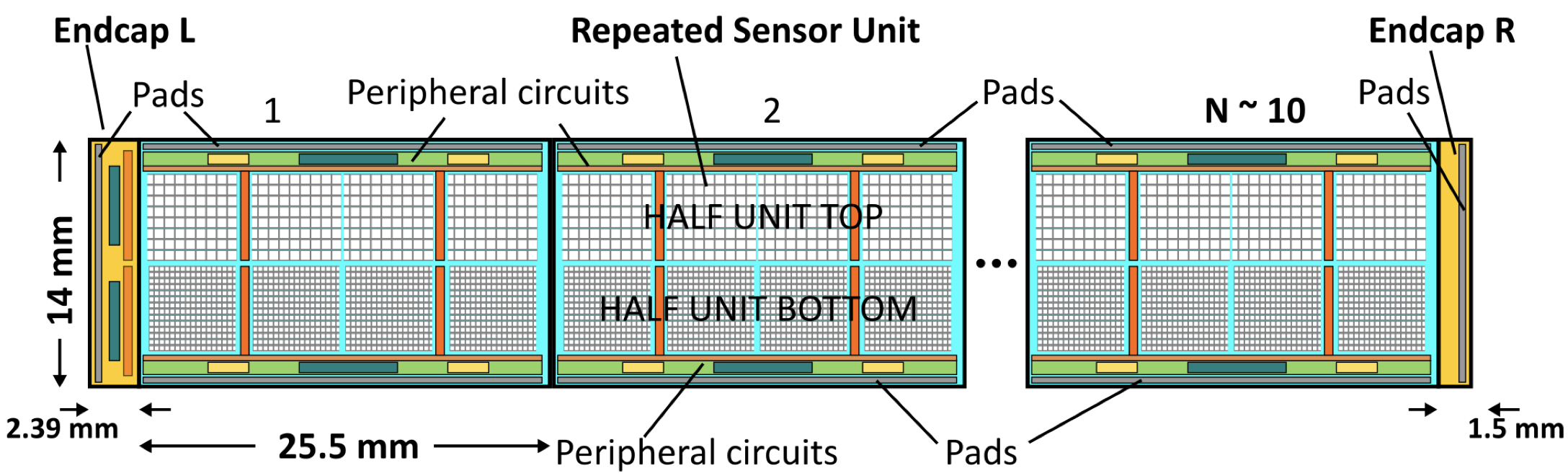}
    \caption{A schematic view of the stitching of the Repeated Sensor Units with typically 10 repetitions.}
    \label{stitch}       
\end{figure}

\section{MLR1 Test Structures}
To qualify the Tower Partners Semiconductor (TPSCo) $65\,$nm CMOS Imaging Sensor (CIS) process for the production of monolithic sensors, an array of test structures has been produced in a 'multi-layer reticle' test production (MLR1).
Different pixel test structures, radiation tests, and analog building blocks have been tested.
The sensors are tested in analog (APTS) and digital (DPTS) pixel test structures.
High efficiency is obtained with the $65\,$nm technology.
Analog building blocks, such as band gap or LVDS drivers, were investigated in these test structures.
Characterisation of pixels has been performed successfully.

\section{Summary}
In these proceedings, the successful completion of several steps for the development of wafer-size stitched sensors has been presented.
The ITS3 project has already passed important milestones: bending of MAPS, characterisation of bent sensors, qualification of the $65\,$nm process, and development of mechanical carbon foam  supports.
Major challenges have been individually investigated and solved.
Thus, the project is ready for full scale prototypes.

\bibliography{proceeding}

%
%

\end{document}